\documentclass[twocolumn,twoside,slac_two]{revtex4}
\usepackage{graphicx}
\usepackage{fancyhdr}
\begin{document}

\title{Faraday rotation in the MOJAVE blazars: Connection with gamma-ray studies}

%

\author{T. Hovatta}
\affiliation{Cahill Center for Astronomy \& Astrophysics, California Institute of Technology, 1200 E. California Blvd, Pasadena, CA 91125, USA}
\affiliation{Department of Physics, Purdue University, 525 Northwestern Ave. West Lafayette, IN 47907, USA}
\author{M. L. Lister}
\affiliation{Department of Physics, Purdue University, 525 Northwestern Ave. West Lafayette, IN 47907, USA}
\author{M. F. Aller, H. D. Aller}
\affiliation{Department of Astronomy, University of Michigan, 817 Dennison Building, Ann Arbor, MI 48109-1042, USA}
\author{D. C. Homan}
\affiliation{Department of Physics and Astronomy, Denison University, Granville, OH 43023, USA}
\author{Y. Y. Kovalev}
\affiliation{Astro Space Center of Lebedev Physical Institute,
  Profsoyuznaya 84/32, 117997 Moscow, Russia}
\affiliation{Max-Planck-Institut f\"ur Radioastronomie, Auf dem
  H\"ugel 69, 53121 Bonn, Germany}
\author{A. B. Pushkarev}
\affiliation{Pulkovo Observatory, Pulkovskoe Chaussee 65/1, 196140
  St. Petersburg, Russia}
\affiliation{Crimean Astrophysical Observatory, 98409 Nauchny, Crimea, Ukraine}
\affiliation{Max-Planck-Institut f\"ur Radioastronomie, Auf dem
  H\"ugel 69, 53121 Bonn, Germany}
\author{T. Savolainen}
\affiliation{Max-Planck-Institut f\"ur Radioastronomie, Auf dem
  H\"ugel 69, 53121 Bonn, Germany}

\begin{abstract}
We have conducted a survey of Faraday rotation in a sample of 191 compact radio-loud AGNs as part of the 
MOJAVE (Monitoring of Jets in Active galactic nuclei with VLBA Experiments) project. 
The observations were carried out with the VLBA at 8.1, 8.4, 12.1 and 15.3 GHz over 12 
epochs in 2006. We detect sufficiently strong linear 
polarization in 159 out of 211 observations to calculate the rotation measure values, resulting 
in a large enough sample for statistical analysis of the Faraday rotation in blazars.
These Faraday rotation measures can be used to study the intrinsic magnetic field order and 
orientation in parsec-scale blazar jets. Our sample includes 119 sources listed in the 1FGL or 2FGL catalogs 
and we detect rotation measure values in 111 out of 131 maps. Of the 72 sources 
that are not in the gamma-ray catalogs we detect RM in 48 out of 80 maps.
The median RM values of the LAT-detected sources do not differ significantly from the non-LAT-detected sources. 
Nine of the sources in our sample have resolved enough jets to study the transverse Faraday rotation structure, 
and we detect significant transverse rotation measure gradients in four sources. In two of these (3C~273 and 
3C~454.3) there is additional evidence to support helical magnetic field in the parsec-scale jets. The two 
others (0923+392 and 2230+114) require further observations to identify the nature of the gradient.
It is interesting that three of the four sources with significant rotation measure gradients are sources 
that have shown large gamma-ray flares. 

\end{abstract}

\maketitle

\thispagestyle{fancy}


\section{Introduction}
Active Galactic Nuclei (AGN) jets are launched from a rotating black hole or accretion disk 
\cite{blandford77, meier01,vlahakis04,mckinney07}. The acceleration and collimation of the jets 
arises naturally if there is differential rotation in the source which winds the poloidal magnetic field lines 
forming a toroidal magnetic field structure \citep{meier01}.  General relativistic magnetohydrodynamic 
simulations have shown that with this kind of setup it is possible to form stable jets with Lorentz factors 
$\Gamma \sim 10$ \cite{mckinney09}.

One way to study the magnetic fields in AGN jets is by multifrequency polarimetric observations. With the 
resolution of the Very Long Baseline Array (VLBA), we can probe the parsec-scale polarization structure of the jets. 
In the optically thin part of the synchrotron spectrum, the magnetic field is perpendicular to the Electric Vector Position Angle (EVPA). 
Therefore by making polarimetric observations, we can study the magnetic field 
structure of the jets. This is important because we do not know if the magnetic field continues to be toroidally dominated 
parsecs down from the central engine and the acceleration and collimation zone, or if it becomes tangled 
due to interaction with the external medium or re-collimation shocks \cite{marscher08}. 

In radio wavelengths one further complication arises due to Faraday rotation. When polarized waves travel through 
non-relativistic plasma, the right and left hand circularly polarized waves have different phase velocities causing a phase offset 
between the two waves. This is seen as a rotation of the EVPA in the plane of the sky and as a diminished degree of 
polarization. In order to study the intrinsic magnetic field orientation, the effect of Faraday rotation must be removed. 
Faraday rotation can occur both internal and external to the source \cite{burn66}. In the case of internal rotation it is caused by 
the low-energy end of the relativistic particle spectrum or by thermal plasma intermixed with the emitting material.
If the rotation is external to the source, it can occur anywhere between the observer and the source, in our own Galaxy, 
intergalactic medium or in the narrow line region or the sheath of the jet. The effect can be described by a linear dependence between the 
observed EVPA ($\chi_\mathrm{obs}$) and wavelength squared ($\lambda^2$) by 
\begin{equation}\label{eq:RM}
\chi_\mathrm{obs} = \chi_0 + \frac{e^3\lambda^2}{8\pi^2\epsilon_0m^2c^3}\int n_e \mathbf{B} \cdot \mathbf{\mathrm{d}l} = \chi_0 + \mathrm{RM}\lambda^2,
\end{equation}
where $\chi_0$ is the intrinsic EVPA and RM is the rotation measure, related to the electron density $n_e$ in the Faraday rotating material and 
the magnetic field component $\mathbf{B}$ parallel to the line of sight so that the RM is positive when the field is coming towards 
the observer and negative when it is going away from the observer. The RM 
can thus be estimated by observing EVPAs at several frequencies.

\section{Observations and data reduction}
The MOJAVE (Monitoring of Jets in Active galactic nuclei with VLBA Experiments) survey monitors the parsec scale structure and evolution of 
a sample of $\sim$ 300 AGN in total intensity and 
polarization with the VLBA at 15\,GHz \cite{lister09}. In 2006 the observations were expanded to include multifrequency polarimetric 
observations of 191 sources at 8.1, 8.4, 12.1 and 15.4\,GHz. Twenty of the sources were observed twice during the year.

The initial data reduction and calibration were performed following the standard procedures described 
in the AIPS cookbook\footnote{http://www.aips.nrao.edu}. All the frequency bands were 
treated separately throughout the data reduction process. The imaging and self-calibration
were done in a largely automated way using the Difmap package \cite{shepherd97}. 
For more details see \cite{lister09} for the standard data reduction and imaging process and 
\cite{lister05} for the calibration of the polarization data.

The {\it (u,v)} plane coverage was matched to be similar at all the bands and all the images were 
restored with the beam size of our lowest frequency, 8.1\,GHz. The absolute EVPA calibration was 
performed using the VLA/VLBA polarization calibration 
database\footnote{http://www.aoc.nrao.edu/$\sim$smyers/calibration/}, MOJAVE 15\,GHz observations, 
University of Michigan Radio Observatory (UMRAO) 8 and 14.5\,GHz observations, and D-term calibration 
method \cite{gomez02,lister05}, resulting in accuracy of $3^\circ$ at 15\,GHz, $2^\circ$ at 12.1\,GHz and $4^\circ$ at the 8\,GHz bands. 
The images of different bands were aligned using Gaussian modelfit components in the 
optically thin part of the jet and 2-D cross correlation. 
The contribution of Faraday rotation in our own Galaxy was removed from the maps by using a smoothed 
Galactic Faraday rotation image \cite{taylor09}. For more details, Hovatta et al. 2012 \cite{hovatta12} (hereafter H2012)

\section{Results}
\begin{figure*}[t]
\centering
\includegraphics[width=135mm]{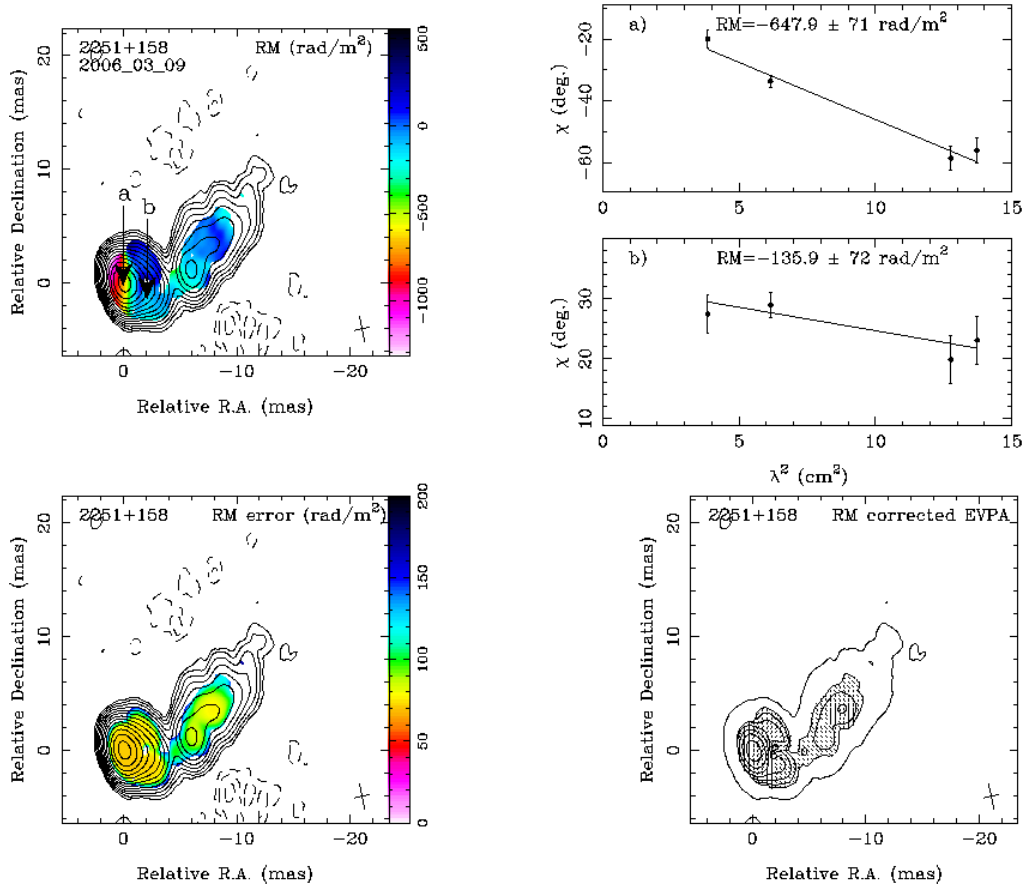} 
\caption{RM map of 3C~454.3 (top left) with $\lambda^2$-fits of two locations in the jet (top right); RM error map (bottom left). Color scale shows the amount of RM in units rad~m$^{-2}$. The contours in the RM and error maps are total intensity contours at 15\,GHz with a peak at 9.1 Jy/beam in successive integer powers of two times the lowest contour 1.6 mJy/beam which is three times the rms level of the total intensity image. A single negative $I$ contour equal to the base contour is also plotted. In the bottom right plot we show the RM corrected electric vectors. The contours include the lowest total intensity contour and polarization contours  with a peak at 0.3 Jy/beam in increasing power of two times the lowest contour 1.6 mJy/beam which is three times the average rms in the $Q$ and $U$ images. Figure from H2012.} \label{fig:3C454}
\end{figure*}

The Faraday rotation measure (RM) can be determined from the slope of the EVPA against $\lambda^2$ as shown by Eq.~\ref{eq:RM}. 
An example of a RM map in the source 3C~454.3 is shown in Fig.~\ref{fig:3C454}, along with EVPA fits in two locations 
of the source. The first one coincides with the VLBA core of the source, taken as the optically thick base of the VLBA jet. 
At 15\,GHz this corresponds to the $\tau = 1$ surface of the jet where the source becomes visible at 15\,GHz. It is 
typically parsecs downstream from the actual black hole, and in the case of 3C~454.3, estimated to be about 18~pc from the 
central engine at 43\,GHz \cite{jorstad10}. 

We find the absolute RM values in our sample to range from zero to a few thousand in the observer's frame. More than 80\% of our 
sources have median absolute RM of less than 400 rad~m$^{-2}$, which would rotate the 15\,GHz EVPAs by 
less than 10 degrees. This means that in large surveys like MOJAVE which are done at frequencies at or above 
15\,GHz, the observed EVPA distributions should reflect the true distributions even if the RM is not accounted for. 
However, our sample includes sources with RMs up to 1000 rad~m$^{-2}$ in the jet components so that when 
studying individual sources, especially at frequencies below 15\,GHz, the amount of Faraday rotation needs to be taken into 
account. 

The situation is more complex in the optically thick cores where we find the RM to vary on short time scales in our own observations and 
when comparing to previous studies of the same sources \cite[e.g.][]{zavala03,zavala04}. This is likely due to multiple 
polarized components blending within the finite beam size or alternatively, if the rotation is internal to the jet, due to 
changes in the particle density or magnetic field strength when new components emerge from the core. Because our 
maps are restored with the 8\,GHz beam, our angular resolution is not good enough to resolve these components and 
higher frequency observations are needed to study the true RMs in the core regions of blazars. For a thorough discussion 
of the RM distributions and variability, see H2012.

\subsection{Comparison to LAT detections}
\begin{figure*}[t]
\centering
\includegraphics[angle=-90, width=135mm]{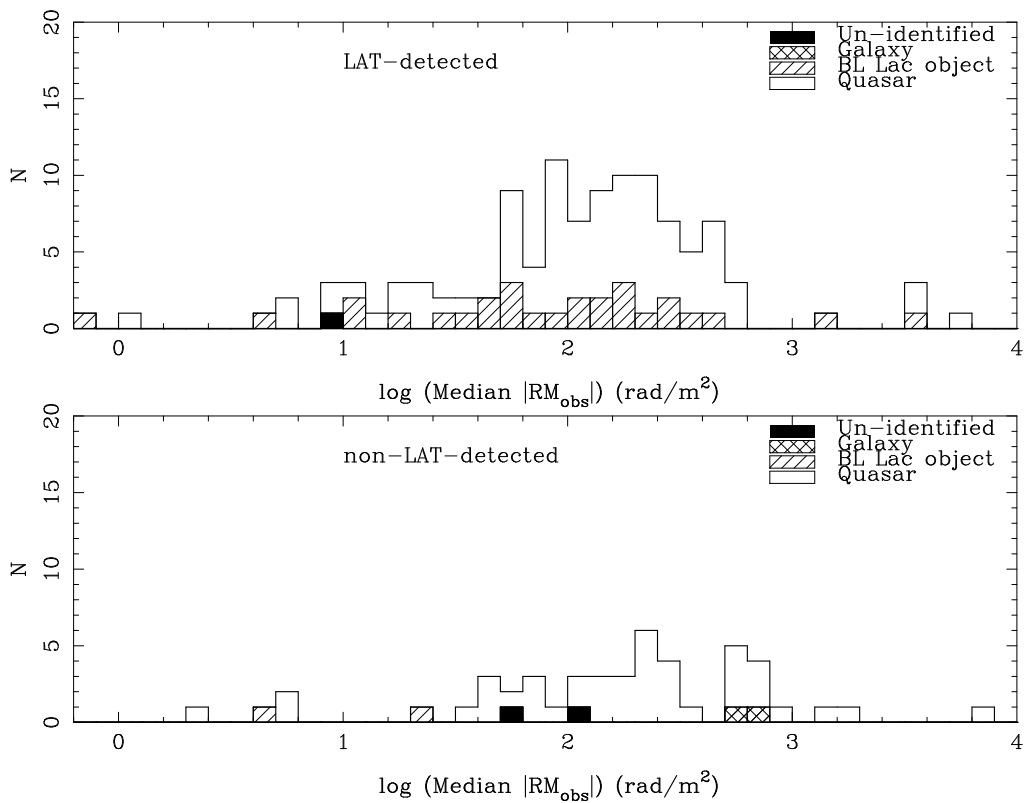}
\caption{Distribution of the median absolute RM in the observer's frame in the LAT-detected sources (top) and non-LAT-detected source (bottom).}\label{fig:dist}
\end{figure*}

Our sample includes 119 sources that are listed in the 1FGL or 2FGL catalogs \cite{abdo10,abdo11}. Twelve of these 
LAT-detected sources were observed twice during the year resulting in 131 maps.
We detect significant polarization in 111 of these cases to be able to calculate the RM maps. The median absolute RM in the 
observer's frame for these cases is 127 rad~m$^{-2}$. Of the 72 non-LAT-detected sources in our sample (8 were observed twice 
resulting in 80 maps), we detect RM only in 48 cases with a median absolute RM of 197 rad~m$^{-2}$ in the observer's frame.
The distributions separated into Flat-Spectrum Radio Quasars (FSRQs), BL Lacs, radio galaxies, and optically un-identified sources are shown in Fig.~\ref{fig:dist}. 
According to the non-parametric Kolmogorov-Smirnov (K-S) test the distributions do not differ significantly (p-value 0.12). 
According to a Welche's t-test the medians of the distributions do not differ significantly. 

The large difference in the fraction of sources where RM is detected in the LAT-detected sources (85\%) compared to the non-LAT-detected
sources (60\%) is due to a positive correlation between linearly polarized flux density and gamma-ray photon flux (Kadler et al. in prep). Thus, sources 
detected by the LAT are more likely to have higher polarized flux density and enable us to detect significant polarization for RM calculation.
A similar trend between polarization and gamma-ray detection is seen at 5\,GHz for the VIPS sample \cite{linford12}.
We note that our observations are not simultaneous with the LAT detections but based on analysis of relation between depolarization and 
RM in the jets, the majority of our RM observations can be explained with a random external screen, in which case the 
RM is not expected to change on time scales of few years (H2012).

\subsection{Transverse RM gradients}
In H2012, we performed detailed simulations of the error in polarization and RM, concentrating on the 
effects of finite beam width and noise in the detection of transverse RM gradients. Such a gradient, where the RM changes 
along a transverse cut over the jet, could be a signature of a helical magnetic field in the jet \cite[e.g.][]{blandford93}. The first 
detection of a transverse RM gradient was reported in the FSRQ 3C~273 \cite{asada02}. Since then there has been several claims of transverse RM 
gradients in various sources \cite[e.g.][]{asada08,mahmud09,contopoulos09} but the issue has remained controversial due to the 
difficulty in assessing the significance of these gradients \cite{taylor10}. Our detailed simulations show that noise in the polarization images 
has a large effect on the detectability of the gradients and the jet needs to be preferably at least two beams wide in polarization with a gradient of 3$\sigma$ in 
significance to call it reliable. Based on these criteria, we find transverse RM gradients in four of our sources, 3C~273, 3C~454.3, 
2230+114, and 0923+392. 

In 3C~273 the gradient is for the first time seen to change sign along the transverse slice. We believe that the difference 
to previous observations where the gradient was significant but the RM values always positive \cite{asada02,asada08b,zavala05} is 
due to different part of the jet being illuminated in our observations. In 3C~273 we also find skewed total intensity, polarization and 
spectral index profiles across the jets, which is a further indication of a helical magnetic field \cite{clausen11}. Similar signatures 
are seen in 3C~454.3 when our observations are combined with other VLBA measurements (Zamaninasab et al. 2012, in prep.).
In both of these two sources, we also detect optically thin jet components with inverse depolarization where the polarization at 
8\,GHz is higher than at 15\,GHz. This is opposite to the typical Faraday depolarization in optically thin jets \cite{burn66} and is 
very difficult to explain with external Faraday rotation models. Instead, the depolarization behavior can be explained with 
internal Faraday rotation (in either helical or loosely tangled  random magnetic field) in which case the rotation is caused by thermal electrons 
intermixed with the emitting plasma or the low-energy end of the particle spectrum \cite{homan12}. Internal 
Faraday rotation is further supported by fast variations on time scales of 3 months seen in the jet RM of these two sources 
(H2012). 

In 0923+392 the gradient is confined to a region near the edge of the jet where the jet is also thought to be bending so that 
it may be caused by interaction with the external medium. In 2230+114 the gradient is visible only in a small part of the jet 
and therefore more sensitive observations are required to confirm the gradient and to be able to model it in detail. 

\section{Discussion and conclusions}
We have conducted a survey of Faraday rotation in 191 sources within the MOJAVE project at 8.1, 8.4, 12.1 and 15.4\,GHz 
with the VLBA. Our sample includes 119 sources detected by the LAT but we do not find a direct relationship between 
the LAT detections and our non-simultaneous Faraday rotation measures.
Despite the lack of direct correlation, the Faraday rotation observations are essential for 
understanding the magnetic field structure in the parsec-scale jets of AGN. Only by resolving the jets is it possible to 
study the line-of-sight magnetic field configuration and find signatures of ordered large-scale magnetic fields, which 
have implications on the jet production models. It is interesting that three out of the four sources where we detect 
significant RM gradients are in sources that are bright gamma-ray emitters. 

The fast RM variability we detect in 3C~273 and 3C~454.3 is hard to explain with external Faraday rotation. Furthermore, the
inverse depolarization we observe in the optically thin jet components in these sources (and the LAT-detected source 1514-241) 
points towards internal Faraday rotation. An additional indication of internal Faraday rotation is non-$\lambda^2$-law behavior 
seen in the optically thin jet components of seven sources, only one of which, the CSS galaxy 0429+415, is not LAT-detected.
This could help us constrain the low-energy end of the particle spectrum which will affect 
estimates of the kinetic luminosity of the jets \cite{celotti93}. Additionally, we will be able to make estimates of the particle density and 
magnetic field strength in the jets.


%




\bigskip 
\begin{acknowledgments}
The MOJAVE project is supported under National Science Foundation grant AST-
0807860 and NASA Fermi grant NNX08AV67G. Work at UMRAO has been supported 
by a series of grants from the NSF and NASA and by funds
for operation from the University of Michigan. T. Hovatta was supported in part by the
Jenny and Antti Wihuri foundation. D. Homan was funded by National Science Foundation grant AST-0707693. 
Part of this work was done when T. Savolainen and Y. Y. Kovalev were research 
fellows of the Alexander von Humboldt Foundation. Y. Y. Kovalev was supported in part by the Russian Foundation 
for Basic Research (grant 11-02-00368) and the Dynasty Foundation. The VLBA is a facility of the 
National Science Foundation operated by the National
Radio Astronomy Observatory under cooperative agreement with Associated Universities,
Inc. This research has made use of NASA's Astrophysics Data System, and the NASA/IPAC Extragalactic Database (NED). 
The latter is operated by the Jet Propulsion Laboratory, California Institute of Technology, under contract with the National 
Aeronautics and Space Administration.
\end{acknowledgments}

\bigskip 

\end{document}